\documentstyle[sprocl]{article}

\input{psfig}

\def\be{\begin{equation}}
\def\ee{\end{equation}}
\def\bea{\begin{eqnarray}}
\def\eea{\end{eqnarray}}

\begin{document}
\hfill{\vbox{\hbox{IFT--P.008/98}\vspace{.3cm}
                \hbox{hep-ph/9801250}}
                }
\title{Limits on Anomalous Couplings from Higgs Boson
       Production at the Tevatron\footnote{Invited talk 
       at the International Workshop on Physics {\sl Beyond the Standard Model: from
Theory to Experiment}, based on the work by F.\ de Campos, S.\ F.\ Novaes, and
M.\ C.\ Gonzalez-Garcia in Ref.\cite{our}.}}

\author{ M.\ C.\ Gonzalez--Garcia}

\address{Instituto de F\'{\i}sica Te\'orica, 
             Universidade Estadual Paulista \\   
             Rua Pamplona 145,
             01405--900 S\~ao Paulo, Brazil\\ and \\ 
          Instituto de F\'{\i}sica Corpuscular - IFIC/CSIC,
             Departament de F\'{\i}sica Te\`orica \\
             Universitat de Val\`encia, 46100 Burjassot, 
             Val\`encia, Spain}


\maketitle
\abstracts{We estimate the attainable limits on the coefficients of
dimension--6 operators from the analysis of Higgs boson
phenomenology, in the framework of a $SU_L(2) \times U_Y(1)$
gauge--invariant effective Lagrangian.  Our results, based on the
data sample already collected by the collaborations at Fermilab
Tevatron, show that the coefficients of Higgs--vector boson
couplings can be determined with unprecedented accuracy. Assuming
that the coefficients of all ``blind'' operators are of the same
magnitude, we are also able to impose more  restrictive bounds on
the anomalous vector--boson triple couplings than the present
limit from double gauge boson production at the Tevatron
collider.}

In the framework of effective Lagrangians respecting the local
$SU_L(2) \times U_Y(1)$ symmetry linearly realized,  the
modifications of the couplings of the Higgs field ($H$) to the
vector gauge bosons ($V$) are related to the anomalous triple
vector boson vertex \cite{linear,drghm,hisz,hsz}. 
The general set of dimension--6
operators involving gauge bosons and the Higgs field, respecting
local $SU_L(2) \times U_Y(1)$ symmetry, and $C$ and $P$
conserving  contains eleven operators \cite{linear,drghm}.
Some of these operators either affect only the Higgs
self--interactions or contribute to the gauge boson two--point
functions at tree level and can be strongly constrained from low
energy physics \cite{drghm,hisz}. The remaining five ``blind''
operators can be written as \cite{linear,drghm,hisz},
\begin{eqnarray}
{\cal L}_{\mbox{eff}}  & & = \sum_i \frac{f_i}{\Lambda^2} {\cal O}_i 
= \frac{1}{\Lambda^2} \Bigl[ 
f_{WWW}\, Tr[\hat{W}_{\mu \nu}\hat{W}^{\nu\rho}\hat{W}_{\rho}^{\mu}] 
+ f_W (D_{\mu} \Phi)^{\dagger} \hat{W}^{\mu \nu} (D_{\nu} \Phi) \nonumber \\
& & + f_B (D_{\mu} \Phi)^{\dagger} \hat{B}^{\mu \nu} (D_{\nu} \Phi) 
+ f_{WW} \Phi^{\dagger} \hat{W}_{\mu \nu} \hat{W}^{\mu \nu} \Phi  
+ f_{BB} \Phi^{\dagger} \hat{B}_{\mu \nu} \hat{B}^{\mu \nu} \Phi 
  \Bigr] \label{lagrangian} 
\end{eqnarray}
where $\Phi$ is the Higgs field doublet, and 
$\hat{B}_{\mu\nu} = i (g'/2) B_{\mu \nu} \;\;\hat{W}_{\mu \nu} = i (g/2)
\sigma^a W^a_{\mu \nu}$
with $B_{\mu \nu}$ and $ W^a_{\mu \nu}$ being the field strength
tensors of the $U(1)$ and $SU(2)$ gauge fields respectively. 

In the unitary gauge, the operators ${\cal O}_{W}$ and ${\cal
O}_{B}$ give rise to both anomalous Higgs--gauge boson couplings
and to new triple and quartic self--couplings amongst the gauge
bosons, while the operator ${\cal O}_{WWW}$ solely modifies the
gauge boson self--interactions \cite{hsz}. 
Searches for deviations on the couplings $WWV$ ($V = \gamma, Z$)
have been carried out at different colliders and recent results
include the ones by CDF \cite{CDF}, and D\O
~Collaborations \cite{D0}. Forthcoming perspectives on this
search at LEP II CERN Collider \cite{lep2,snow}, and at upgraded
Tevatron Collider \cite{tevatron} were also reported. 

The operators  ${\cal O}_{WW}$ and ${\cal O}_{BB}$ only affect
$HVV$ couplings, like $HWW$, $HZZ$, $H\gamma\gamma$ and
$HZ\gamma$, since their contribution to the $WW\gamma$ and $WWZ$
tree--point couplings can be completely absorbed in the
redefinition of the SM fields and gauge couplings. Therefore, one
cannot obtain any constraint on these couplings from the study of
anomalous trilinear gauge boson couplings. These anomalous
couplings were extensevely studied in electron--positron collisions
\cite{hsz,epem}.

We consider here Higgs production at the Fermilab
Tevatron collider with its subsequent decay into two photons
\cite{h:smw}. This channel in the SM occurs at one--loop level
and it is quite small, but due to the new interactions
(\ref{lagrangian}), it can be enhanced and even become dominant.
 We study the associated  $HV$ process (\ref{assoc}) and 
  the vector boson fusion process (\ref{fusion}),  
\begin{eqnarray}
p \bar{p} &\to & q \bar{q}^\prime \to W/Z (\to f  \bar{f}^\prime) +
H (\to \gamma \gamma) \; ,  
\label{assoc}        \\
p \bar{p} &\to & q \bar{q}^\prime WW (ZZ) \to j + j +
H (\to \gamma \gamma) \; , 
\label{fusion}
\end{eqnarray}
taking into account the 100 pb$^{-1}$ of integrated luminosity
already collected by the Fermilab Tevatron collaborations.
We focus on the signatures $\ell \nu \gamma \gamma$, 
($\ell = e, \mu$), and  $j \, j \, \gamma \gamma$,
coming from the reactions (\ref{assoc}) and (\ref{fusion}). 

Recently, D\O ~Collaboration has presented their results for the
search of high invariant--mass photon pairs in $p\bar{p}
\rightarrow \gamma\gamma j j$ events \cite{D0:jjgg}. We show,
based on their results, that it may be possible to obtain  a
significant indirect limit on anomalous $WWV$ coupling under the
assumption that the coefficients of  the ``blind'' effective
operators contributing to the Higgs--vector boson couplings are
of the same magnitude.  It is also possible to restrict the
operators that involve just Higgs boson couplings, $HVV$, and
therefore can not be bounded by the $W^+W^-$ production at LEPII.

We have included in our calculations all SM (QCD plus
electroweak), and anomalous contributions that lead to these
final states. The SM one-loop contributions to the $H\gamma\gamma$
and $H Z\gamma$ vertices were introduced through the use of the 
effective operators with the corresponding form factors in the coupling. 
Neither the narrow--width approximation for the Higgs boson 
contributions, nor the effective $W$ boson approximation were employed. We
consistently included the effect of all interferences between the
anomalous signature and the SM background. A total of 42 (32) SM
(anomalous) Feynman diagrams are involved in the subprocesses of
$\ell \nu \gamma \gamma$ \cite{lngg} for each leptonic flavor,
while 1928 (236) participate in $j \, j \, \gamma \gamma$
signature \cite{jjgg}. The SM Feynman diagrams 
were generated by Madgraph
\cite{madgraph} in the framework of Helas \cite{helas}. The
anomalous contributions arising from the Lagrangian
(\ref{lagrangian}) were implemented in Fortran routines and were
included accordingly. We have used the MRS (G) \cite{mrs} set of
proton structure functions with the scale $Q^2=\hat{s}$.

The cuts applied on the final state particles are similar to
those used  by the experimental collaborations \cite{CDF,D0}.
In particular when studying the $\gamma\gamma j j $ final state
we have closely followed the results recently presented by D\O
~Collaboration  \cite{D0:jjgg}.
We also assumed an invariant--mass resolution for the
two--photons of  $\Delta M_{\gamma\gamma}/ M_{\gamma\gamma} =
0.15/ \sqrt{M_{\gamma\gamma}} \oplus 0.007$ \cite{h:smw}. Both
signal and background were integrated over an invariant--mass bin
of $\pm 2 \Delta M_{\gamma\gamma}$ centered around $M_H$.

The signature of the $j\, j \, \gamma\gamma$ process receives
contributions from both associated production and $WW/ZZ$ fusion.
We isolate the majority
of events due to associated production, and the corresponding
background, by integrating over a bin centered on the $W$ or $Z$
mass, which is equivalent to the invariant mass cut listed above. 

After imposing all the cuts, we get a reduction on the signal
event rate which depends on the Higgs mass.  For the $jj
\gamma\gamma$ final state the geometrical acceptance and
background rejection cuts account for a reduction factor of 15\%
for $M_H=60$ GeV rising to 25\% for $M_H=160$ GeV. We also
include in our analysis the particle identification and trigger
efficiencies which vary from 40\% to 70\% per particle lepton or
photon \cite{D0}. For the $jj\gamma\gamma$  ($\ell
\nu\gamma\gamma$) final state we estimate the total effect of
these  efficiencies to be 35\% (30\%). We therefore obtain an
overall efficiency for the $jj\gamma\gamma$ final state of 5.5\%
to 9\% for $M_H = 60$--$160$ GeV in agreement with the results of
Ref.\ \cite{D0:jjgg}.

Dominant backgrounds are due to missidentification when
a jet fakes a photon  what has been estimated to occur with a
probability  of a few times $10^{-4}$ \cite{D0}. Although this
probability is small, it becomes the main source of  background
for the $j j \gamma\gamma$ final state because of the very large
multijet cross section. In Ref.\ \cite{D0:jjgg} this background
is estimated to lead to $3.5\pm 1.3$ events with invariant mass
$M_{\gamma\gamma}>60 $ GeV and it has been consistently included
in our derivation of the attainable limits. 

In the $l\nu \gamma\gamma$ channel the dominant fake background
is $W\gamma j$ channel, when the jet mimics a photon. We
estimated the contribution of this channel to yield $N_{back}<
0.01$ events \cite{D0} at 95\% CL. We have also  estimated the
various QCD fake backgrounds  such as $jjj$, $jj \gamma$ and
$j\gamma\gamma$ with the jet faking a photon and/or electron plus
fake missing missing, which are  to be negligible.  

The coupling $H\gamma\gamma$ derived from (\ref{lagrangian}) involves $f_{WW}$
and $f_{BB}$ \cite{hsz}.  In consequence, the anomalous signature $f\bar f
\gamma\gamma$ is only possible when those couplings are not vanishing. The
couplings $f_B$ and $f_W$, on the other hand, affect the production mechanisms
for the Higgs boson. In what follows, we present our results for three
different scenarios of the anomalous coefficients:  $(i)$ Suppressed $VVV$
couplings compared to the $H\gamma\gamma$ vertex: $f_{BB,WW} = f \gg f_{B,W}$; 
$(ii)$ all coupling with the same magnitude and sign: $f_{BB,WW,B,W} = f$;
$(iii)$ all coupling with the same magnitude but different relative sign:
$f_{BB,WW} = f = - f_{B,W}$.

In order to establish the attainable bounds on the coefficients,
we imposed an upper limit on the number of signal events based on
Poisson statistics. For the $jj\gamma\gamma$
final state we use the results from Ref.\ \cite{D0:jjgg}, where
no event has been reported in the $100$ pb$^{-1}$ sample. For the
other cases, the limit on the number of signal events was
conservatively obtained assuming that the number of observed
events coincides with the expected background. 

Table \ref{tab} shows the range of $f/\Lambda^2$ that can be
excluded at 95\% CL with the present Tevatron luminosity in the
scenario $(i)$. We should remind that this scenario will not be
restricted by LEP II data on $W^+W^-$ production since there is
no trilinear vector boson couplings involved. As seen in the
table, the best limits are obtained for the $j j \gamma\gamma$
final state and they  are more restrictive than the ones coming
from $e^+ e^- \to \gamma\gamma\gamma$ or $b \bar{b}\gamma$ at LEP
II \cite{epem}.

\begin{table}[t]
\caption{Allowed range of $f/\Lambda^2$ in TeV$^{-2}$ at 95\% CL,
assuming the scenario $(i)$ ($f_{BB}=f_{WW}\gg f_B, f_W$) for the
different final states, and for different Higgs boson masses for
an integrated luminosity of 100 pb$^{-1}$.\label{tab}}
\vspace{0.2cm}
\begin{center}
\footnotesize
\begin{tabular}{||c@{\hspace{-0.2cm}}c||l@{\hspace{-0.03cm}}
|l@{\hspace{-0.03cm}}|l@{\hspace{-0.03cm}}|l@{\hspace{-0.03cm}}||}
\hline
\hline
 $M_H$/GeV &        & 100 & 150 & 200 & 250  \\
\hline
\hline
$\ell \, \nu \,\gamma\, \gamma$    & RunI &
                             ($-41$ --- 74) & 
                             ($-83$ --- 113) &
                             ($<-200$ --- $>200$) &  
                              $<-200$--- $>200$ \\
&  RunII &
                             ($-13$ --- 36) & 
                             ($-22$ --- 46) &
                             ($-57$ --- 135) &  
                             ($-195$ --- $>200$)   \\  
&  TeV33 &
                             ($-3.8$ --- 8) & 
                             ($-4.8$ --- 20) &
                             ($-28$ --- 60) &  
                             ($-45$ --- 83)  
                             \\
\hline
$j \, j \, \gamma \, \gamma$ &   RunI &
                             ($-20$ --- 49) & 
                             ($-26$ --- 64) &
                             ($-96$ ---$>100 $) &  
                             ($<-100$ --- $>100$)   
                             \\
&  RunII &
                             ($-8.4$ --- 26) & 
                             ($-11$ --- 31) &
                             ($-36$ --- 81) &  
                             ($-64$ --- $>100$) \\ 
&  TeV33 &
                             ($-4.2$ --- 6.5) & 
                             ($-4.5$ --- 12) &
                             ($-19$ --- 40) &  
                             ($-28$ --- 51) \\
\hline\hline
\end{tabular}
\end{center}
\end{table}

For the scenarios $(ii)$ and $(iii)$, the limits derived from our
study lead to constraints on the triple gauge boson coupling
parameters.  The most general parametrization for the $WWV$
vertex can be found in Ref.\ \cite{classical}. When only the
operators (\ref{lagrangian}) are considered, it contains three
independent parameters. If it is further assumed that $f_B=f_W$,
only two free parameters remain, which are usually chosen as
$\Delta\kappa_\gamma$ and $\lambda_\gamma$. This is usually
quoted in the literature as the HISZ scenario \cite{hisz}. 

\begin{figure}[t]
\begin{center}
\psfig{figure=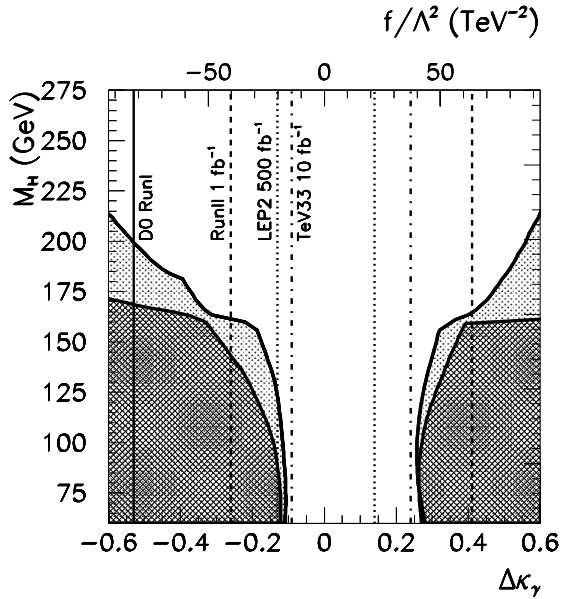,height=2.5in}
\end{center}
\caption{Excluded region in the $\Delta\kappa_\gamma \times M_H$
plane for an integrated luminosity of 100 pb$^{-1}$, and for
scenarios $(ii)$ (clear shadow) and $(iii)$ (dark shadow).  The
present and future bounds on $\Delta\kappa_\gamma$ are also shown
(see text for details).\label{fig:2} }
\end{figure}

Since we are assuming  $f_B=f_W$ our results can be compared to
the derived limits from triple gauge boson studies in the HISZ
scenario. In Fig.\ \ref{fig:2}, we show the region in
the $\Delta \kappa_\gamma \times M_H$ that can be excluded
through the analysis of the present Tevatron data, accumulated in
Run I, with an integrated luminosity of 100 pb$^{-1}$
\cite{D0:jjgg}, for scenarios $(ii)$ and $(iii)$.

For the sake of comparison, we also show in Fig.\ \ref{fig:2} the
best available  experimental limit on $\Delta \kappa_\gamma$
\cite{D0} and the expected bounds, from double gauge
boson production, from an updated Tevatron Run II, with 1
fb$^{-1}$, and TeV33 with 10 fb$^{-1}$ \cite{tevatron}, and from
LEP II operating at 190 GeV with an integrated luminosity of 500
fb$^{-1}$ \cite{snow}. In all cases the results were obtained
assuming the HISZ scenario. We can see that, for $M_H \leq
200 [170]$ GeV, the limit that can be established at 95\% CL from
the Higgs production analysis for scenario $(ii)$ [$(iii)$],
based on the present Tevatron luminosity is tighter than the
present limit coming from gauge boson production. 

When the same analysis  is performed for the upgraded Tevatron, a
more severe restriction on the coefficient of the anomalous
operators is obtained. For instance, from $p\bar{p} \to j \, j \,
\gamma \gamma$, in scenario $(ii)$ we get, for $M_H=150$ GeV: 
For RunII with 1 fb$^{-1}$  
$- 9 < f < 25$  $(-0.06 < \Delta\kappa_\gamma < 0.16)$; 
For TeV33 with 10 fb$^{-1}$ $- 4 < f < 15$  $(- 0.03 <\Delta\kappa_\gamma <
0.1)$. 



\section*{References}
\begin{thebibliography}{99}

\bibitem{our}  F.\ de Campos, M.\ C.\ Gonzalez-Garcia and 
S.\ F.\ Novaes, Phys.\ Rev.\ Lett.\  {\bf 79}, 5210 (1997).

\bibitem{classical}  K.\ Hagiwara, H.\ Hikasa, R.\ D.\ Peccei and D.\
Zeppenfeld, Nucl.\ Phys.\ {\bf B282}, 253 (1987).
 
\bibitem{linear} C.\ J.\ C.\ Burguess and H.\ J.\ Schnitzer,
Nucl.\ Phys.\ {\bf B228}, 464 (1983); C.\ N.\ Leung, S.\ T.\ Love
and S.\ Rao, Z.\ Phys.\ {\bf 31}, 433 (1986);  W.\ Buchm\"uller
and D.\ Wyler, Nucl.\ Phys.\ {\bf B268}, 621 (1986).

\bibitem{drghm} A.\ De Rujula, M.\ B.\ Gavela, P.\ Hernandez and
E.\ Masso, Nucl.\ Phys.\ {\bf B384}, 3 (1992); A.\ De Rujula, 
M.\ B.\ Gavela, O.\ Pene and F.\ J.\ Vegas, 
Nucl.\ Phys.\ {\bf B357}, 311 (1991).

\bibitem{hisz} K.\ Hagiwara, S.\ Ishihara, R.\ Szalapski and D.\
Zeppenfeld,  Phys.\ Lett.\ {\bf B283}, 353 (1992); {\it idem},
Phys.\ Rev.\ {\bf D48}, 2182 (1993); K.\ Hagiwara, T.\ Hatsukano,
S.\ Ishihara and  R.\ Szalapski, Nucl.\ Phys.\ {\bf B496}, 66 (1997).

\bibitem{CDF} F.\ Abe {\it et al.}, CDF Collaboration, Phys.\
Rev.\ Lett.\ {\bf 74}, 1936 (1995); {\it idem} {\bf 74}, 1941 
(1995); {\it idem} {\bf 75}, 1017 (1995); {\it idem} {\bf 78}, 4536 (1997).

\bibitem{D0} S.\ Abachi {\it et al.}, D\O  ~Collaboration, Phys.\
Rev.\ Lett.\ {\bf 75}, 1023 (1995);  {\it idem} {\bf 75}, 1028 (1995); 
{\it idem} {\bf 75}, 1034 (1995); {\it idem} {\bf 77}, 3303 (1996);  
{\it idem} {\bf 78}, 3634 (1997); {\it idem} {\bf 78}, 3640 (1997);
B.\ Abbott {\it et al.}, D\O  ~Collaboration, Phys.\ 
Rev.\ Lett.\ {\bf 79}, 1441 (1997).

\bibitem{lep2} For a review see:  Z.\ Ajaltuoni {\it et al.},
``Triple Gauge Boson Couplings", in {\it Proc. of the CERN
Workshop on LEPII Physics}, edited by G.\ Altarelli, T.\
Sj\"ostrand, and F.\ Zwirner, CERN 96--01, Vol. 1, p. 525 (1996).

\bibitem{snow} T.\ Barklow {\it et al.}, Summary of the Snowmass
Subgroup on Anomalous Gauge  Boson Couplings, to appear in the
{\it Proceedings of the 1996 DPF/DPB Summer Study on New
Directions in High-Energy Physics}, June 25 --- July 12 (1996),
Snowmass, CO, USA.

\bibitem{tevatron} D.\ Amidei {\it et al.}, {\it Future Electroweak
Physics at the Fermilab Tevatron: Report of the TeV--2000 Study
Group}, preprint FERMILAB-PUB-96-082 (1996).

\bibitem{hsz} K.\ Hagiwara, R.\ Szalapski and D.\ Zeppenfeld, 
Phys.\ Lett.\ {\bf B318}, 155 (1993).

\bibitem{D0:jjgg} B.\ Abbott {\it et al.}, D\O  ~Collaboration, 
FERMILAB-CONF-97/325-E, contribution to the Lepton-Photon Conference, 
Hamburg, July 1997.

\bibitem{epem} G.\ J.\ Gounaris, J.\ Layssac and F.\ M.\ Renard, 
Z.\ Phys.\ {\bf C65}, 245 (1995); G.\ J.\ Gounaris, F.\ M.\ Renard 
and N.\ D.\ Vlachos, Nucl.\ Phys.\ {\bf B459}, 51 (1996);
S.\ M.\ Lietti, S.\ F.\ Novaes and R.\
Rosenfeld, Phys.\ Rev.\ {\bf D54}, 3266 (1996); F.\ de Campos,
S.\ M.\ Lietti, S.\ F.\ Novaes and R.\ Rosenfeld, Phys.\ Lett.\
{\bf B389}, 93 (1996).

\bibitem{h:smw} A.\ Stange, W.\ Marciano, and S.\ Willenbrock, Phys.\
Rev.\ {\bf D49}, 1354 (1994); Phys.\ Rev.\ {\bf D50}, 4491 (1994).

\bibitem{lngg} R.\ Keiss, Z.\ Kunszt, and W.\ J.\ Stirling, Phys.\ Lett.\ 
{\bf B253}, 269 (1991); J.\ Ohnemus and W.\ J.\ Stirling,
Phys.\ Rev.\ {\bf D47}, 336 (1992);  U.\ Baur, T.\ Han, N.\ Kauer,
R.\ Sobey, and D.\ Zeppenfeld, Phys.\ Rev.\ D{\bf 56}, 140 (1997). 

\bibitem{jjgg} V.\ Barger, T.\ Han , D.\ Zeppenfeld, and J.\ Ohnemus, Phys.\
Rev.\ {\bf D41}, 2782 (1990).


\bibitem{madgraph} T.\ Stelzer and W.\ F.\ Long, Comput.\ Phys.\
Commun.\  {\bf 81}, 357 (1994).

\bibitem{helas} H.\ Murayama, I.\ Watanabe and K.\ Hagiwara, KEK
report 91-11.

\bibitem {mrs} A.\ D.\ Martin, W.\ J.\ Stirling, R.\ G.\ Roberts
Phys.\ Lett.\ {\bf B354}, 155 (1995).

\end{thebibliography}
\end{document}